\documentclass[conference]{IEEEtran}

\usepackage{graphicx}
\usepackage{psfrag}
\usepackage[cmex10]{amsmath}
\usepackage{amssymb}
\usepackage{ifthen}
\usepackage[amsmath,thmmarks]{ntheorem}
\usepackage{my_theorems}
\usepackage{epsfig}

\DeclareMathOperator{\Tr}{T}

\DeclareMathOperator{\He}{H}

\DeclareMathOperator*{\minimize}{minimize}

\DeclareMathAlphabet{\mathbit}{OML}{cmr}{bx}{it}

\newcommand{\B}[1]{\mathbit{#1}}
\newcommand{\R}[1]{\mathrm{#1}}

\newcommand{\id}{\mathbf{I}}
\newcommand{\zero}{\mathbf{0}}

\newcommand{\sigman}{\sigma_{\eta}^2}

\newcommand{\Ptx}{P_{\R{Tx}}}

\begin{document}

\title{On the Convexity of the MSE Region of Single-Antenna Users}

\author{\IEEEauthorblockN{Raphael Hunger and Michael Joham}
\IEEEauthorblockA{Associate Institute for Signal Processing,
Technische Universit\"at M\"unchen, 80290 Munich, Germany\\
Telephone: +49 89 289-28508, Fax: +49 89 289-28504, Email: \texttt{\{hunger,joham\}@tum.de}
}}

\maketitle

\begin{abstract}
  We prove convexity of the sum-power constrained 
  \emph{mean square error} (MSE) region in case of
  two single-antenna users communicating with a multi-antenna base station.
  Due to the MSE duality this holds both for the vector broadcast channel and 
  the dual multiple
  access channel. Increasing the number of users to more than two, 
  we show by
  means of a simple counter-example that the resulting MSE region 
  is not necessarily convex any longer, even under the assumption of single-antenna users.
  In conjunction with our former observation that the two user MSE region is
  not necessarily convex for two multi-antenna users, this extends and corrects
  the hitherto existing notion of the MSE region geometry.
  
\end{abstract}

\IEEEpeerreviewmaketitle

\section{Introduction}

  Up to now, only few contributions on the geometrical structure of the 
  mean square error region exist.
  In~\cite{schubert05n}, the authors show that the multi-user MIMO
  MSE region is convex under \emph{fixed} transmit and receive beamforming
  vectors both for linear and nonlinear preprocessing.
  Obviously, a larger set of MSE tuples can be achieved 
  by means of \emph{adaptive} transmit and receive beamformers. 
  For this extended setup only the two user case has been investigated
  so far. Utilizing matrix inequalities of matrix-convex functions,
  the authors in \cite{JB03tsft} prove that the two user
  multi-antenna MSE region cannot exhibit a nonconvex dent
  between two feasible MSE points connected by a line segment
  with $-45^\circ$ slope. From this observation, they claim that the MSE region
  is convex. For convexity, however, all possible slopes would have to
  be checked. As a matter of fact, a channel realization exhibiting
  a nonconvex MSE region with two multi-antenna users 
  has been observed in \cite{HuJoUTb07}
  disproving the convexity theorem in \cite{JB03tsft}.
  A multi-carrier system where several single-antenna users communicate
  with a single-antenna base station has been investigated in~\cite{WBM07uobo}.
  There, the complementary MSE region of parallel broadcast channels
  is shown to be not necessarily convex. Since the system under
  consideration in~\cite{WBM07uobo} 
can be recast into a \emph{block diagonal} MIMO broadcast channel,
  the authors of \cite{WBM07uobo} conclude that the two user multi-antenna MSE region
  cannot be convex in general which again contradicts the theorem in \cite{JB03tsft}.
  So far, no distinct statements on convexity of the MSE region depending
  on the number of users and antennas per user are available in case of 
  adaptive transmit and receive beamformers. 
  
  Some applications for which the geometry of the MSE region is of interest
  are for example the stream priorization according to buffer states
  or queue states by means of the weighted sum-MSE minimization,
  cf.~\cite{Codreanu_2006}. Here, suboptimum transmit and receive filters
  are derived by repeatedly switching between the downlink and the dual uplink
  in combination with a geometric program solver for a reasonable
  power allocation. Balancing is
  considered in~\cite{MeJoHuUt06} where the weights of a weighted
  sum-MSE minimization are adapted until certain MSE ratios hold.
  Exploiting the relationship between the derivative of the 
  the mutual information and the minimum mean square error, 
  Christensen et al.\ tackle the weighted sum-rate maximization
  utilizing results from a weighted sum-MSE minimization, 
  see~\cite{Christensen_2007}.
  However, convexity 
  of the MSE region is the crucial point for the proper functionality 
  of above applications since nonconvexity may for example 
  prevent convergence of iterative algorithms. 
  Finally, the MSE $\varepsilon$ achieved with MMSE receivers
  is tightly related to the maximum SINR via 
  \begin{equation}
     \mathrm{SINR} = \frac{1}{\varepsilon}-1,\\
  \end{equation}
  and hence, also to the data rate $R$ via the simple relation
  \begin{equation}
    R  = -\log_2 \varepsilon.
  \end{equation}
  Summing up, all this clearly motivates a detailed investigation.

  In this paper, we extend the
  hitherto existing notion of the MSE region geometry.
  The single antenna case with two users is covered
  in Section~\ref{sec:two_users} whereas statements on the convexity
  of the MSE region for three or more single-antenna users are presented
  in Section~\ref{sec:three_users}. Finally, a conjecture on the
  convexity of the
  multi-antenna two user case is given in Section~\ref{sec:conjecture},
  and detailed proofs for the 
  theorems and corollaries are attached in 
  Appendices~\ref{sec:first_proof}--\ref{sec:third_proof}
  for the sake of readability.

\section{Convexity for Two Single-Antenna Users }
\label{sec:two_users}
\begin{figure}[!t]
    \centering
    \psfrag{1}{$1$}
    \psfrag{e2}{$\varepsilon_2$}
    \psfrag{e1}{$\varepsilon_1$}
    \psfrag{e2min}{\hspace{-3mm}$\varepsilon_{\text{min},2}$}
    \psfrag{e1min}{$\varepsilon_{\text{min},1}$}
    \psfrag{g}{{\small $\varepsilon_2=g(\varepsilon_1)$}}
    \includegraphics[width=2in]{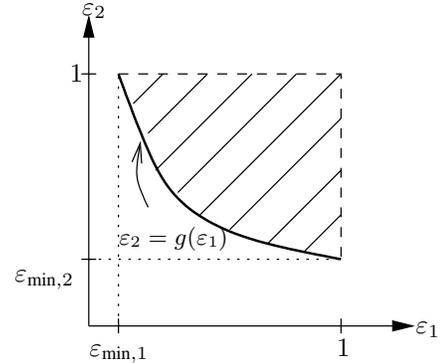}
    \caption{MSE $\varepsilon_2$ of user $2$ depending on MSE $\varepsilon_1$ of
    user $1$.}
    \label{fig:function_g_figure}
\end{figure}

In this section we present statements on the geometry of the MSE region
of two single-antenna users. For this setup, convexity can always
be shown:
\Theorem{MSE_convexity}{section}
{
  The MSE region of two single-antenna users is convex
  both in the multiple-access channel and in the vector broadcast channel.
}  
\begin{IEEEproof}
  See Appendix~\ref{sec:first_proof}.
\end{IEEEproof}
For the most important part of the boundary of the two user MSE region
(see Fig.~\ref{fig:function_g_figure})
there is a functional relationship 
\begin{equation}
  \varepsilon_2=g(\varepsilon_1)
\end{equation}
between the two users' MSEs $\varepsilon_1$ and $\varepsilon_2$.
If the channel vectors $\B{h}_1$ and $\B{h}_2$ describing the
transmission from both users to the base station in the dual MAC
are not colinear, the function $g$ is \emph{strictly} convex, otherwise,
it is affine:
\Corollary{curvature_zero_corollary}{section}
{
  The function $g:\varepsilon_1\mapsto \varepsilon_2=g(\varepsilon_1)$  describing
  the efficient set of the MSE region is strictly convex if 
  $\B{h}_1$ and $\B{h}_2$ are not colinear. Otherwise, $g$ is affine.
}

\begin{IEEEproof}
  See Appendix~\ref{sec:second_proof}.
\end{IEEEproof}

\section{Nonconvexity Example for More than Two Single-Antenna Users}
\label{sec:three_users}

Although the MSE region is convex for two single-antenna users,
this property may get lost when adding an additional 
user, even if he is equipped with only a single antenna:
\Theorem{three_user_MSE_region}{section}
{
  The three user MSE region of both the vector broadcast channel and
  the multiple-access channel is not necessarily convex.
}
\begin{IEEEproof}
  For the proof, we present a simple example 
  in Appendix~\ref{sec:third_proof} where the line segment
  connecting two feasible MSE triples lies \emph{outside} the MSE region.
  A further confirmation of Theorem~\ref{three_user_MSE_region}
  results from the observation, that the weighted sum-MSE
  minimization has more than one local minimum, see Appendix~\ref{sec:third_proof}.
\end{IEEEproof}
Nonconvexity implies for example that not every point of the MSE 
efficient set
can be achieved by means of the weighted sum-MSE minimization technique.
Balancing approaches based on the weighted sum-MSE minimization algorithm
hence may fail to achieve the desired MSE ratios, cf.~\cite{HuJoUTb07}. Instead they are
prone to oscillations. 
The following theorem covers the case when (three or) more than three users are 
present in the system:
\Theorem{four_user_mse_region}{three_user_MSE_region}
{
  The MSE region of more than two users may be nonconvex both
  in the vector broadcast channel and in the multiple-access channel.
}
\begin{IEEEproof}
The three user case has already been shown in
Theorem~\ref{three_user_MSE_region}.
For more than three users, the MSE region is a multi-dimensional
manifold. However, setting the powers of those users to $p_4=\ldots=p_K=0$, the intersection of this
manifold with the $K-3$ hyperplane(s) $p_i=0, \ i\in\{4,\ldots,K\}$,
is again a three-dimensional manifold which may have the same geometry
as the manifold of the three user case. Hence, the MSE region may be
nonconvex for more than three users as well.
\end{IEEEproof}

\section{Conjecture on the Convexity of the Multi-Antenna Two User Case}
\label{sec:conjecture}

A counter-example to convexity of the MSE region
when multi-antenna users are involved has been shown in 
\cite{HuJoUTb07},
where two users each equipped with two antennas communicate with
a multi-antenna base-station. 
Similarly, the multi-carrier single-antenna system in \cite{WBM07uobo} 
can be recast into a multi-antenna MIMO broadcast channel system where again
nonconvexity was observed.
Following the idea in the proof of Theorem~\ref{four_user_mse_region},
the MSE region of two or more than two 
users may be nonconvex as soon as two multi-antenna users
are present.
Proving convexity for the case of one
single-antenna user and one multi-antenna user turns out to be
difficult since a parametrization of the lower left boundary 
of the feasible MSE region is not known, points on this boundary
are obtained by limits of iterative algorithms. Nonetheless, extensive
simulation results bring us to the \emph{conjecture} that the MSE region
of one single antenna user and one multi-antenna user is convex.


\appendices

\section{Proof of Theorem~\ref{MSE_convexity}}
\label{sec:first_proof}

  Because of the MSE duality between the vector BC~\cite{YuCioffi} and the MAC in
 \cite{HuJoUt07,MeJoHuUt06,schubert05o},
  it suffices to prove convexity in the MAC which is easier to handle.
  Fig.~\ref{fig:function_g_figure} shows the basic characteristics of the two user
  MSE region for single-antenna users. Here, the MSEs $\varepsilon_1$
  and $\varepsilon_2$ of both users are upper bounded by~$1$ since 
  MMSE receivers are assumed. Allowing for other receiver types
  does not bring any reasonable gain since only MSE-pairs where at least 
  one entry
  may lie above~$1$ would arise. Under the assumption of MMSE receivers,
  the right part of the
  boundary of the MSE region is obtained when user one 
  does not transmit any data to the 
  base station at all and user two varies its transmit power from
  zero to $\Ptx$. Similarly, the upper part of the boundary 
  is reached when user two does not transmit at all whereas user one
  varies its transmit power from zero to $\Ptx$.
  Evidently, the most interesting part of the boundary is the lower left
  one, where the sum of both transmit powers equals
  the maximum available power $\Ptx$. MSE pairs lying on this boundary
  feature the functional relationship 
  $\varepsilon_2=g(\varepsilon_1)$, where the domain and the image of $g$
  are the sets $[\varepsilon_{\text{min},1},1]$ and
  $[\varepsilon_{\text{min},2},1]$, respectively. When less than the
  total transmit power $\Ptx$ is consumed, points are achieved that
  are element of the 
  interior of the MSE region. As a conclusion, 
  convexity of the set of feasible MSE points corresponds to convexity of the
  function~$g$ relating the MSE $\varepsilon_1$ of user one to the MSE
  $\varepsilon_2$ of user two on the lower left boundary of the MSE region. 
  In the following, we show that 
  \begin{equation}
    \frac{\partial^2 \varepsilon_2}{\partial \varepsilon_1^2}=
    \frac{\partial^2 g(\varepsilon_1)}{\partial \varepsilon_1^2}  \geq 0
    \label{second_derivative}    
  \end{equation}
  holds which immediately implies convexity of $g$.

  Unfortunately, a direct functional relationship  
  between $\varepsilon_2$ and $\varepsilon_1$ 
  is not available. Instead, the two MSEs $\varepsilon_1$ and $\varepsilon_2$
  are parametrized by the transmit power of one of them, for example
  by the transmit power $p\in\mathbb{D}=[0,\Ptx]$ of user one:
  \begin{equation*}
    \varepsilon_1 = f_1(p), \quad \varepsilon_2 = f_2(p).
  \end{equation*}
  We can conclude that user two has to transmit with power
  $\Ptx-p$ in order to utilize the complete power budget.
  In conjunction with MMSE receivers, the mean square error of user one reads as
  \begin{equation}
    \varepsilon_1 = f_1(p) = 1-p\B{h}_1^{\He}\B{X}^{-1}(p)\B{h}_1 > 0,
    \label{MSE_user_1}   
  \end{equation}
  with the positive definite covariance matrix of the received signal
  \begin{equation}
    \B{X}(p) = \sigman\id_N+p\B{h}_1\B{h}_1^{\He}+(\Ptx-p)\B{h}_2\B{h}_2^{\He}
    \label{receive_covariance}  
  \end{equation}
  and $\sigman>0$ represents the variance of the noise at every antenna element.
  Similarly, the MSE of user two is denoted by
  \begin{equation}
    \varepsilon_2 = f_2(p)= 1 - (\Ptx\!-\!p)\B{h}_2^{\He}\B{X}^{-1}(p)\B{h}_2 >0.
    \label{MSE_user_2}     
  \end{equation}
  Combining (\ref{MSE_user_1}), (\ref{MSE_user_2}), and (\ref{receive_covariance}),
  the function $f_1$ turns out to be strictly monotonically decreasing in~$p$,
  i.e.,
  \begin{equation}
    \dot{\varepsilon}_1:=\frac{\partial f_1(p)}{\partial p} < 0 \quad 
    \forall p\in\mathbb{D},
    \label{f_1_decreasing}  
  \end{equation}
  whereas $f_2$ is strictly monotonically increasing in $p$:
  \begin{equation}
    \dot{\varepsilon}_2:=\frac{\partial f_2(p)}{\partial p} > 0 
    \quad \forall p\in\mathbb{D}.
    \label{f_2_increasing}  
  \end{equation}
  From (\ref{f_1_decreasing}) and (\ref{f_2_increasing}), 
  pseudo-convexity of $g$ already follows.
  Before validating~(\ref{second_derivative}), we compute the first derivative:
  \begin{equation}
    \frac{\partial g(\varepsilon_1)}{\partial \varepsilon_1} = 
    \left.\frac{\frac{\partial f_2(p)}{\partial p}}
         {\frac{\partial f_1(p)}{\partial
	 p}}\right|_{p=f_1^{-1}(\varepsilon_1)}.
  \label{g_derivative}	 
  \end{equation}
  Note that $f_1^{-1}(\varepsilon_1)$ denotes the inverse function of $f_1$
  which exists due to~(\ref{f_1_decreasing}). Differentiating
  (\ref{g_derivative}) again with respect to $\varepsilon_1$ yields
  \begin{equation}
    \begin{split}
       \frac{\partial^2 g(\varepsilon_2)}{\partial \varepsilon_1^2} & = 
       \frac{\partial}{\partial \varepsilon_1}
       \Bigg(\left.\frac{\frac{\partial f_2(p)}{\partial p}}
         {\frac{\partial f_1(p)}{\partial p}}\right|_{p=f_1^{-1}(\varepsilon_1)}
	 \Bigg) \\
	 & = \left.\Bigg(\frac{\partial}{\partial p}\
	     \frac{\frac{\partial f_2(p)}{\partial p}}
	          {\frac{\partial f_1(p)}{\partial p}}
	       \Bigg)\right|_{p=f_1^{-1}(\varepsilon_1)}
	       \cdot \frac{\partial f_1^{-1}(\varepsilon_1)}{\partial\varepsilon_1}\\
         & = \left.\frac{\ddot{\varepsilon}_2\dot{\varepsilon}_1 - 
	 \ddot{\varepsilon}_1\dot{\varepsilon}_2}
	     {\left(\dot{\varepsilon}_1\right)^2}\right|_{p=f_1^{-1}(\varepsilon_1)}
	     \cdot \frac{1}{\dot{\varepsilon}_1|_{p=f_1^{-1}(\varepsilon_1)}}\\
         & = \left.\frac{\ddot{\varepsilon}_2\dot{\varepsilon}_1 - 
	 \ddot{\varepsilon}_1\dot{\varepsilon}_2}
        {\left(\dot{\varepsilon}_1\right)^3}\right|_{p=f_1^{-1}(\varepsilon_1)}.
    \end{split}
    \label{second_derivative_final}
  \end{equation}
  Since $f^{-1}_1$ maps from $[\varepsilon_{\text{min},1},1]$ to $\mathbb{D}$,
  and since $\dot{\varepsilon}_1<0$ holds $\forall p\in\mathbb{D}$, 
  the function $g$ is convex iff 
  [see (\ref{second_derivative_final}) and cond.\ (\ref{second_derivative})]
  \begin{equation}
    \ddot{\varepsilon}_2\dot{\varepsilon}_1 - 
	 \ddot{\varepsilon}_1\dot{\varepsilon}_2 \leq 0
	 \Leftrightarrow g \ \text{is convex}.
   \label{convexity_condition}	 
  \end{equation}
  For notational brevity, we introduce the two substitutions
  \begin{equation}
    a_{i,j} = \B{h}_i^{\He}\B{X}^{-1}(p)\B{h}_j \quad \text{and} \quad
    b_{i,j} = \B{h}_i^{\He}\B{X}^{-2}(p)\B{h}_j,
    \label{substitutions}    
  \end{equation}
  which satisfy $a_{i,j}=a_{j,i}^*$ and $b_{i,j}=b_{j,i}^*$.
  Making use of 
  \begin{displaymath}
    \frac{\partial \B{X}^{-1}(p)}{\partial p} = 
    - \B{X}^{-1}(p)\frac{\partial \B{X}(p)}{\partial p}\B{X}^{-1}(p), 
  \end{displaymath}
  the first derivatives with respect to~$p$ in (\ref{f_1_decreasing}) and (\ref{f_2_increasing})
  can be shown to equal
  \begin{equation}
    \label{diff_expressions_1}    
    \begin{split}
      \dot{\varepsilon}_1 & = -\sigman b_{1,1} - \Ptx|a_{1,2}|^2,\\
      \dot{\varepsilon}_2 & = +\sigman b_{2,2} + \Ptx|a_{1,2}|^2,
    \end{split}
  \end{equation}
  respectively. Differentiating (\ref{diff_expressions_1}) again w.r.t.\ $p$, we obtain
  \begin{equation}
    \begin{split}
       \ddot{\varepsilon}_1 = 
            2\sigman [a_{1,1}b_{1,1} - \Re\{a_{1,2}b_{2,1}\}]
	    +2\Ptx|a_{1,2}|^2(a_{1,1}-a_{2,2}),\\
       \ddot{\varepsilon}_2 = 
            2\sigman [a_{2,2}b_{2,2} - \Re\{a_{2,1}b_{1,2}\}]
            +2\Ptx|a_{1,2}|^2(a_{2,2}-a_{1,1}).    
    \end{split}
    \nonumber
  \end{equation}
  Inserting (\ref{diff_expressions_1}) 
  and the last two equations into~(\ref{convexity_condition}) 
  and applying $\Re\{a_{2,1}b_{1,2}\}=\Re\{a_{1,2}b_{2,1}\}$
  results in
  \begin{equation}
    \begin{split}
      & \ddot{\varepsilon}_2\dot{\varepsilon}_1 - 
	 \ddot{\varepsilon}_1\dot{\varepsilon}_2 = \\
      & \ \ \ 2\sigman\Ptx|a_{1,2}|^2
      \left[2\Re\{a_{1,2}b_{2,1}\}-a_{2,2}b_{1,1}-a_{1,1}b_{2,2}\right]\\
      & \ \ \ + 2\sigma_{\eta}^4b_{1,1}\left(\Re\{a_{1,2}b_{2,1}\}
        -a_{1,1}b_{2,2}\right)\\
      & \ \ \ + 2\sigma_{\eta}^4b_{2,2}\left(\Re\{a_{1,2}b_{2,1}\}
        -a_{2,2}b_{1,1}\right).
    \end{split}
    \label{derivative_difference}
  \end{equation}
  In order to prove that (\ref{derivative_difference}) is not positive
  to fulfill the convexity requirement in~(\ref{convexity_condition}),
  we will reveal that all three summands in~(\ref{derivative_difference}) 
  are not positive. 
  
  For the first summand, this turns out to be very 
  easy: Noticing that $a_{i,i}>0$ and $b_{i,i}>0$, the first summand
  in (\ref{derivative_difference}) is nonpositive if
  \begin{equation}
    4\Re^2\{a_{2,1}b_{1,2}\}\leq (a_{2,2}b_{1,1}+a_{1,1}b_{2,2})^2.
    \label{first_comparison}    
  \end{equation}
  Clearly, we can upper bound the real part by the magnitude and apply
  the \emph{Cauchy-Schwarz}-inequality with (\ref{substitutions}) to bound the magnitude:
  \begin{equation}
    4\Re^2\{a_{2,1}b_{1,2}\}\leq 4|a_{2,1}b_{1,2}|^2
    \leq 4a_{2,2}a_{1,1}b_{1,1}b_{2,2}.
    \label{two_upper_bounds}
  \end{equation}
  Validating the inequality 
  \begin{equation}
    \begin{split}
      (a_{2,2}b_{1,1}+a_{1,1}b_{2,2})^2 & \geq 4a_{2,2}a_{1,1}b_{1,1}b_{2,2}\\
   \Leftrightarrow   (a_{2,2}b_{1,1}-a_{1,1}b_{2,2})^2& \geq 0
    \end{split}
    \nonumber    
  \end{equation}
  leads in conjunction with (\ref{two_upper_bounds}) to the conclusion
  that (\ref{first_comparison}) is fulfilled, i.e.,
  the first summand in~(\ref{derivative_difference}) is 
  nonpositive.
  
  Nonpositivity of the second summand in (\ref{derivative_difference})
  is resembled by the inequality
  \begin{equation}
    \Re\left\{\frac{a_{1,2}}{a_{1,1}}\frac{b_{2,1}}{b_{2,2}}\right\}\leq 1.
    \label{second_comparison}
  \end{equation}
  To prove (\ref{second_comparison}) we explicitly have to exploit
  the structure of $\B{X}(p)$ in (\ref{receive_covariance})
  which makes the proof longer than the one for the first summand.
  Interestingly, the real part operator in (\ref{second_comparison}) is
  redundant as its argument turns out to be real-valued.
  Applying the matrix inversion lemma several times,
  we get
  \begin{equation}
    \frac{a_{1,2}}{a_{1,1}}=\frac{\sigman\B{h}_1^{\He}\B{h}_2}
    {\sigman\|\B{h}_1\|_2^2+d(\Ptx\!-\!p)},
    \label{fraction_1}    
  \end{equation}
  with the substitution
  \begin{equation}
    d=\|\B{h}_1\|_2^2\|\B{h}_2\|_2^2\!-\!|\B{h}_1^{\He}\B{h}_2|^2\geq 0.
    \label{d_substitution}    
  \end{equation}
  Applying several times the matrix inversion lemma
  as for the first fraction, 
  the second fraction in~(\ref{second_comparison}) can be expressed as
  \begin{equation}
    \frac{b_{2,1}}{b_{2,2}}=\frac{\B{h}_2^{\He}\B{h}_1
    \left[\sigma_{\eta}^4-p(\Ptx\!-\!p)d\right]}
    {\sigma_{\eta}^4\|\B{h}_2\|_2^2+dp(2\sigman+p\|\B{h}_1\|_2^2)}.
   \label{fraction_2}    
  \end{equation}
  Multiplying (\ref{fraction_1}) by (\ref{fraction_2}) yields the real-valued
  expression
  \begin{equation*}
    \frac{b_{2,1}a_{1,2}}{a_{1,1}b_{2,2}} = 
    \frac{\sigma_{\eta}^6 |\B{h}_1^{\He}\B{h}_2|^2 - c_1}
      {\sigma_{\eta}^6\|\B{h}_1\|_2^2\|\B{h}_2\|_2^2 + c_2} \ \in\mathbb{R}
  \end{equation*}
  with the two substitutions
  \begin{equation*}
    \begin{split}
      c_1 & = \sigman|\B{h}_1^{\He}\B{h}_2|^2 pd(\Ptx-p) \geq 0,\\
      c_2 & = \left[\sigman\|\B{h}_1\|_2^2+d(\Ptx-p)\right]dp\left(2\sigman+p\|\B{h}_1\|_2^2\right)\\
      & \quad + \sigma_\eta^4\|\B{h}_2\|_2^2d(\Ptx-p) \geq 0.
    \end{split}
  \end{equation*}
  Since both $c_1$ and $c_2$ are nonnegative, we find
  \begin{equation*}
    \frac{b_{2,1}a_{1,2}}{a_{1,1}b_{2,2}} \leq 
    \frac{\sigma_{\eta}^6 |\B{h}_1^{\He}\B{h}_2|^2}
      {\sigma_{\eta}^6\|\B{h}_1\|_2^2\|\B{h}_2\|_2^2}
  \end{equation*}
  as an upper bound from which (\ref{second_comparison}) 
  directly follows due to the
  \emph{Cauchy-Schwarz}-inequality. Thus, the nonpositivity of the second
  summand in (\ref{derivative_difference}) is proven.
  
  Finally, the nonpositivity of the third summand in (\ref{derivative_difference}) 
  is shown by the same reasoning as for the second summand:
  \begin{equation}
    \Re\left\{\frac{a_{1,2}}{a_{2,2}}\frac{b_{2,1}}{b_{1,1}}\right\}\leq 1    
  \end{equation}
  is deduced from 
  \begin{equation*}
    \frac{b_{2,1}a_{1,2}}{a_{2,2}b_{1,1}} = 
    \frac{\sigma_{\eta}^6 |\B{h}_1^{\He}\B{h}_2|^2 - d_1}
      {\sigma_{\eta}^6\|\B{h}_1\|_2^2\|\B{h}_2\|_2^2 + d_2} \ \in\mathbb{R},
  \end{equation*}
  where $d_i$ follows from $c_i$ by interchanging indices and powers:
  \begin{equation*}
    \begin{split}
       d_1 & = c_1,\\
       d_2 & = \left(\sigman\|\B{h}_2\|_2^2+dp\right)d(\Ptx-p)
           \left[2\sigman+(\Ptx-p)\|\B{h}_2\|_2^2\right]\\
      & \quad + \sigma_\eta^4\|\B{h}_1\|_2^2dp \geq 0.
    \end{split}
  \end{equation*}
  As all three summands in (\ref{derivative_difference}) are nonpositive,
  the inequality in (\ref{convexity_condition}) is satisfied and the 
  proof for the convexity of the MSE region is complete.
  

\section{Proof of Corollary~\ref{curvature_zero_corollary}}
\label{sec:second_proof}

  If the inequality in (\ref{convexity_condition}) is strict
  for all $p\in\mathbb{D}$,
  $g$ is strictly convex. Excluding equality in 
  (\ref{convexity_condition}) therefore ensures that $g$ is
  strictly convex. The difference in
  (\ref{derivative_difference}) is zero if and only if
  all three summands are zero since each summand is nonpositive. 
  In order to let
  the first summand vanish, the \emph{Cauchy-Schwarz}-inequality
  in (\ref{two_upper_bounds}) has to be fulfilled with equality. To this end, 
  $\B{h}_1$ and $\B{h}_2$ have to be colinear which also fulfills
  (\ref{first_comparison}) with equality.
  If both channel vectors are colinear, $d=0$ results from
  (\ref{d_substitution}) and the variables $c_1$, $c_2$, $d_1$, and
  $d_2$ are zero as well. Obviously, (\ref{second_comparison})
  holds with equality and the last two summands in (\ref{derivative_difference})
  vanish. Thus, we have shown that if the two channel vectors
  $\B{h}_1$ and $\B{h}_2$ are not colinear, then the function $g$ is
  strictly convex. Additionally, if both vectors are colinear,
  $g$ has curvature zero for all powers $p\in\mathbb{D}$.
  As a consequence, $g$ is affine.
  In the latter case, we have the relationship
  \begin{equation}
      g(\varepsilon_1) =-\varepsilon_1\frac{|\alpha|^2\!+\!|\alpha|^2\gamma\|\B{h}_1\|_2^2}
          {1+|\alpha|^2\gamma\|\B{h}_1\|_2^2}
	  + 1 + \frac{|\alpha|^2}{1+|\alpha|^2\gamma\|\B{h}_1\|_2^2},
  \end{equation}
  where $\gamma=\Ptx/\sigman$ denotes the transmit SNR, 
  $\B{h}_2 = \alpha\B{h}_1$,
  and $\varepsilon_1\in [\varepsilon_{\text{min},1},1]$ with
  \begin{equation}
    \varepsilon_{\text{min},1} = \frac{1}{1+\gamma\|\B{h}_1\|_2^2}. 
  \end{equation}

\section{Proof of Theorem~\ref{three_user_MSE_region}}
\label{sec:third_proof}  

A nonconvex three user MSE region can for example be obtained by the channel
matrix 
\begin{equation}
  \B{H} = [\B{h}_1,\B{h}_2,\B{h}_3] = \left[\begin{array}{ccc}1 & 0 & 1\\0& 1 & 1\end{array}\right]
  \label{joham_a}
\end{equation}
and a transmit power $\Ptx=10$.
In this case, the base station is equipped with $N=2$ antennas,
and the channel vector $\B{h}_3$ is the sum of $\B{h}_1$
and $\B{h}_2$. Note that the base station has fewer antennas than
users are present in the system in this special case. 
Nonconvexity of the MSE region can also be and has been
observed when the channel vectors of all users
are linearly independent ($N\geq K$ must hold then).
If the MSE region was convex, the line segment between every
two feasible MSE triples would have to be a subset of the region.
Moreover, the weighted sum-MSE minimization with arbitrary nonnegative weights 
$\B{w}=[w_1,\ldots,w_K]^{\Tr}\geq \zero_K,\ \B{w}\neq \zero_K$
may have stationary
points fulfilling the KKT conditions
\emph{with only one common value} of the weighted minimization. 
In the following, we show that these conditions are violated 
for the channel in~(\ref{joham_a}).

The weighted sum-MSE minimization reads as
\begin{equation}
    \minimize_{p_1,\ldots,p_K} \sum_{k=1}^K w_k\varepsilon_k
    \quad \text{s.t.:} \ \sum_{k=1}^K p_k \leq \Ptx,\ \ 
     p_k\geq 0\ \forall k,
  \label{wsmm}     
\end{equation}
where $p_k$ is the power with which user~$k$ transmits in the uplink and
the MSE of user~$k$ reads as
\begin{equation}
  \varepsilon_k = 1-p_k\B{h}_k^{\He}\B{X}^{-1}\B{h}_k
\end{equation}
with the received signal covariance matrix
\begin{equation}
  \B{X} = \sigman\id_N + \sum_{\ell=1}^Kp_\ell\B{h}_\ell\B{h}_{\ell}^{\He}.
\end{equation}
The \emph{Lagrangian} function associated to (\ref{wsmm}) reads as
\begin{equation}
  L = \sum_{k=1}^K w_k\varepsilon_k + \lambda\Big(\sum_{k=1}^K p_k-\Ptx\Big)
  - \sum_{k=1}^K \mu_k p_k.
\end{equation} 
Note that the \emph{Lagrangian} multipliers $\lambda$ and $\mu_1,\ldots,\mu_K$
have to be nonnegative real.
If above \emph{Lagrangian} $L$ has stationary points with different values
for $L$, the
underlying MSE region is not convex since more than one 
hyperplane with normal vector $[w_1,\ldots,w_K]^{\Tr}$ locally supporting the
MSE region exists.
The KKT conditions read as
\begin{equation}
  \B{h}_k^{\He}\check{\B{X}}^{-1}(w_k\check{\B{X}}-\check{\B{S}})
   \check{\B{X}}^{-1}\B{h}_k =\check{\lambda} - \check{\mu}_k \ \forall k,
   \label{KKT1}
\end{equation}
\vspace{-1.5mm}
\begin{equation}   
   \check{p}_k \geq 0 \ \forall k,
\end{equation}   
\vspace{-1.5mm}
\begin{equation}   
  \check{p}_k \check{\mu}_k = 0 \ \forall k,
\end{equation}  
\vspace{-1.5mm}
\begin{equation}
  \check{\mu}_k \geq 0 \ \forall k,
\end{equation}
\vspace{-1.5mm}
\begin{equation}
   \sum_{k=1}^K \check{p}_k \leq \Ptx,
\end{equation}   
\vspace{-1.5mm}
\begin{equation}
   \check{\lambda} \Big(\sum_{k=1}^K \check{p}_k -\Ptx\Big)= 0,
\end{equation}
\begin{equation}
  \check{\lambda}\geq 0,
  \label{KKT5}
\end{equation}
with the substitution
\vspace{-2mm}
\begin{equation}
  \B{S}=\sum_{\ell=1}^K w_\ell p_\ell\B{h}_{\ell}\B{h}_\ell^{\He}.
\end{equation}
Note that checked variables $\check{(\cdot)}$ are those which fulfill the KKT conditions.
Assuming a weight vector 
\begin{equation}
  \B{w}=[0.22,0.54,0.24]^{\Tr},
\end{equation}
the weighted sum-MSE minimization (\ref{wsmm}) features \emph{two}
stationary points satisfying the KKT conditions
(\ref{KKT1})--(\ref{KKT5}) for the channel vectors (\ref{joham_a}) and a transmit
power $\Ptx=10$. The first set of primal and dual variables fulfilling the KKTs reads as
\begin{equation}
  \begin{split}
    &\check{\B{p}}{}^{(1)}=[3.6753,\ 6.3247,\ 0]^{\Tr}, \ \
    \check{\lambda}{}^{(1)}=0.0101, \\
    &\check{\mu}_1{}^{(1)}=\check{\mu}_2{}^{(1)}=0, \ \
    \check{\mu}_3{}^{(1)}=0.0266,
  \end{split}  
  \label{KKT_point_1}
\end{equation}
and achieves a weighted sum-MSE 
$\sum_{k=1}^3 w_k \check{\varepsilon}_k{}^{(1)}=0.36078$.
The second set of variables reads as
\begin{equation}
  \begin{split}
    &\check{\B{p}}{}^{(2)}=[0,\ 7.0794,\ 2.9206]^{\Tr},\ \
    \check{\lambda}{}^{(2)}=0.0115, \\
    &\check{\mu}_1{}^{(2)}=0.007, \ \
    \check{\mu}_2{}^{(2)}=\check{\mu}_3{}^{(2)}=0,
  \end{split}  
  \label{KKT_point_2}
\end{equation}
and obtains a slightly larger metric
$\sum_{k=1}^3 w_k \check{\varepsilon}_k{}^{(2)}=0.3828$.
The existence of two KKT points with different values
algebraically proves the nonconvexity of the MSE region. 

\begin{figure}[!t]
  \psfrag{a}{$\varepsilon_1$}
  \psfrag{b}{$\varepsilon_2$}
  \psfrag{c}{$\varepsilon_3$}
  \epsfig{file=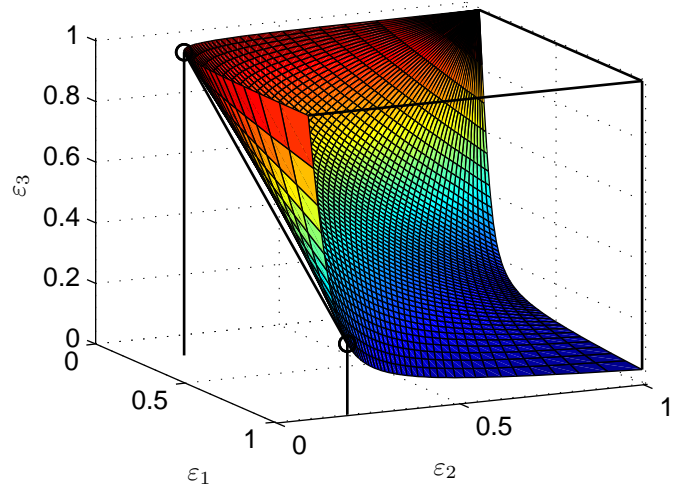}
  \caption{Example of a nonconvex MSE region for $K=3$ users. The line
  segment connecting two feasible points lies \emph{outside} the region.}
  \label{fig:three_users}
\end{figure}

A geometrical proof is shown in Fig.~\ref{fig:three_users},
where the three-user MSE region for the channel in
(\ref{joham_a}) is plotted with $\Ptx=10$.
The two KKT points in (\ref{KKT_point_1})
and (\ref{KKT_point_2}) achieve individual MSEs
\begin{equation}
  \begin{split}
    \check{\boldsymbol{\varepsilon}}{}^{(1)}=
    \big[\check{\varepsilon}_1{}^{(1)},\check{\varepsilon}_2{}^{(1)},
    \check{\varepsilon}_3{}^{(1)}\big]^{\Tr} 
    & = [0.2139, \ 0.1365, \ 1]^{\Tr},\\
    \check{\boldsymbol{\varepsilon}}{}^{(2)}=
    \big[\check{\varepsilon}_1{}^{(2)},\check{\varepsilon}_2{}^{(2)},
    \check{\varepsilon}_3{}^{(2)}\big]^{\Tr} 
    & = [1, \ 0.1977, \ 0.2335]^{\Tr}.
  \end{split}    
\end{equation}
However, the line segment connecting $\check{\boldsymbol{\varepsilon}}{}^{(1)}$
and $\check{\boldsymbol{\varepsilon}}{}^{(2)}$ does not completely
belong to the MSE region, it lies outside the region and touches
the boundary of the MSE region at $\check{\boldsymbol{\varepsilon}}{}^{(1)}$
and $\check{\boldsymbol{\varepsilon}}{}^{(2)}$.
Evidently, the MSE region cannot be convex.

\bibliographystyle{IEEEtran}
\bibliography{IEEEabrv,references}

\begin{thebibliography}{10}
\providecommand{\url}[1]{#1}
\csname url@rmstyle\endcsname
\providecommand{\newblock}{\relax}
\providecommand{\bibinfo}[2]{#2}
\providecommand\BIBentrySTDinterwordspacing{\spaceskip=0pt\relax}
\providecommand\BIBentryALTinterwordstretchfactor{4}
\providecommand\BIBentryALTinterwordspacing{\spaceskip=\fontdimen2\font plus
\BIBentryALTinterwordstretchfactor\fontdimen3\font minus
  \fontdimen4\font\relax}
\providecommand\BIBforeignlanguage[2]{{%
\expandafter\ifx\csname l@#1\endcsname\relax
\typeout{** WARNING: IEEEtran.bst: No hyphenation pattern has been}%
\typeout{** loaded for the language `#1'. Using the pattern for}%
\typeout{** the default language instead.}%
\else
\language=\csname l@#1\endcsname
\fi
#2}}

\bibitem{schubert05n}
S.~Shi and M.~Schubert, ``{Convexity Analysis of the Feasible MSE Region of
  Sum-Power Constrained Multiuser MIMO Systems},'' in \emph{Proc. IEEE
  Internat. Symp. on Personal, Indoor and Mobile Radio Communications (PIMRC),
  Berlin, Germany}, Sept. 2005.

\bibitem{JB03tsft}
E.~Jorswieck and H.~Boche, ``{Transmission Strategies for the MIMO MAC with
  MMSE Receiver: Average MSE Optimization and Achievable Individual MSE
  Region},'' \emph{IEEE Trans. on Signal Processing}, vol.~51, no.~11, pp.
  2872--2881, Nov. 2003, special issue on MIMO wireless communications.

\bibitem{HuJoUTb07}
R.~Hunger, M.~Joham, and W.~Utschick, ``{Efficient MSE Balancing for the
  Multi-User MIMO Downlink},'' in \emph{{Proc.\ 41st Asilomar Conference on
  Signals, Systems, and Computers}}, November 2007.

\bibitem{WBM07uobo}
G.~Wunder, I.~Blau, and T.~Michel, ``{Utility Optimization based on MSE for
  Parallel Broadcast Channels: The Square Utility Optimization based on MSE for
  Parallel Broadcast Channels: The Square Root Law},'' in \emph{{45th Annual
  Allerton Conference on Communication, Control, and Computing}}, Monticello,
  Illinois, September 2007.

\bibitem{Codreanu_2006}
M.~Codreanu, A.~Tolli, M.~Juntti, and M.~Latva-aho, ``{Weighted Sum MSE
  Minimization for MIMO Broadcast Channel},'' in \emph{{17th International
  Symposium on Personal, Indoor, and Mobile Radio Communications (PIMRC)}},
  September 2006.

\bibitem{MeJoHuUt06}
A.~Mezghani, M.~Joham, R.~Hunger, and W.~Utschick, ``Transceiver {D}esign for
  {M}ulti-{U}ser {MIMO} {S}ystems,'' in \emph{Proc.\ ITG/IEEE WSA 2006}, March
  2006.

\bibitem{Christensen_2007}
S.~Christensen, R.~Agarwal, E.~d.~Carvalho, and J.~M. Cioffi, ``{Beamforming
  Design for Weighted Sum-Rate Maximization in MIMO Broadcast Channels},''
  \emph{Submitted to IEEE Transactions on Wireless Communications}.

\bibitem{YuCioffi}
W.~Yu and J.~M. Cioffi, ``{Sum Capacity of Gaussian Vector Broadcast
  Channels},'' \emph{IEEE Trans. Inform. Theory}, vol.~50, pp. 1875--1892,
  September 2004.

\bibitem{HuJoUt07}
R.~Hunger, M.~Joham, and W.~Utschick, ``{On the MSE-Duality of the Broadcast
  Channel and the Multiple Access Channel},'' \emph{Submitted to IEEE
  Transactions on Signal Processing}.

\bibitem{schubert05o}
M.~Schubert, S.~Shi, E.~A. Jorswieck, and H.~Boche, ``{Downlink Sum-MSE
  Transceiver Optimization for Linear Multi-User MIMO Systems},'' in
  \emph{Proc. Asilomar Conf. on Signals, Systems and Computers, Monterey, CA},
  Sept. 2005.

\end{thebibliography}

%

\end{document}